\documentclass{mem}
\usepackage{natbib}\usepackage{txfonts}\usepackage{balance}
\usepackage{graphicx}
\usepackage{txfonts}
\usepackage[a4paper]{hyperref}
\idline{79}{3}
\begin{document}
\def\teff{$T\rm_{eff }$}
\def\kms{$\mathrm {km s}^{-1}$}

\title{
Photometric binaries in fifty  Globular Clusters
}

   \subtitle{}

\author{
A. \,P. \,Milone\inst{1} 
G. \,Piotto\inst{1} 
L. \,R. \,Bedin\inst{2} 
A. \,Sarajedini\inst{4}
          }

  \offprints{A. P. Milone}

\institute{
Dipartimento  di   Astronomia,  Universit\`a  di
  Padova, Vicolo dell'Osservatorio 3, Padova, I-35122, Italy
\and
Space Telescope Science Institute, 3700 San Martin Drive,
Baltimore, MD 21218, USA
\and
Department of Astronomy, University of Florida,
	      211 Bryant Space Science Center, Gainesville, FL 32611, USA
}

\authorrunning{Milone et al.}

\titlerunning{Photometric binaries in fifty GCs}

\abstract{ The HST/ACS Survey  of Galactic globular clusters (GGCs) is
a HST Treasury project  aimed at  obtaining high  precision photometry  in a
large  sample  of  globular  clusters.   The  homogeneous  photometric
catalogs  that  has been  obtained  from  these  data by  Anderson  et
al.  (2008)  represents a  golden  mine  for  a lot  of  astrophysical
studies.

In  this paper we  used the catalog to  analyse the  properties of
MS-MS  binary systems  from  a sample  of  fifty GGCs.   We measured  the
fraction  of binaries  (divided  in different  groups), studied  their
radial distribution  and constrained the mass  ratio distribution.  We
investigated possible  relations between the fraction  of binaries and
the main parameters of their host GGCs.

We found a significant anti-correlation between the binary fraction in
a cluster and its absolute luminosity (mass).

 } \maketitle{}

\section{Introduction}
Knowledge of the binary frequency in Globular Clusters (GCs) 
is of foundamental importance for a lot of astrophisical studies.

Binaries  play an  important  role  in the  dynamical  evolution of  a
clusters. Interactions with hard binaries pump kinetic energy into the
cluster core, slowing the core  collaps and, eventually, causing the core to
reexpand, if the number of binaries is large enough. In general, binaries are a
foundamental ingredient in any dynamical evolution model of a GC.

Exotic stellar  objects, like Blue  Stragglers, cataclismic variables,
millisecond  pulsars and  low  mass  X ray  binaries  are believed  to
represent   evolutionary   stages   of   close  binary   system.   The
determination of  the fraction of  binaries plays a  foundamental step
towards the understanding of the evolution of these peculiar objects.
Furthermore,   binary  stars  introduce   systematic  errors   in  the
determination  of the  main sequence  (MS) fiducial  line and  move it
toward red colors with respect to its correct position.

Finally, a correct determination  of the mass and luminosity functions
requires a correct measure of the fraction of binaries.

Up to now, three main techniques have been used to measure the fraction
ob binaries in GGCs (Hut et al. 1992).

The first  one identifies binaries by measuring  their radial velocity
variation (eg. Latham 1996). This  method relies with the detection of
each individual binary system but, due to the limits in sensibility
of spectroscopy, these studies are possible only for the brightest GGCs
stars.

The second tecnique  is based on the search  for photometric variables
(eg. Mateo  1996). As well  as the previous  one, it is able  to infer
specific properties of each binary system (like the measure of orbital
period, mass  ratio, orbital inclination). Unfortunatly,  it is biased
towards  binaries with  short  periods and  large orbital  inclination.
Moreover these tecniques  have a low discovery efficency  and are very
expensive in terms of telescope time because it is necessary to repeat
measures in time.

A thirth  approach, that  is based  on the analysis  of the  number of
stars  located  on the  red  side  of the main sequence (MS)  ridge
line (MSRL)  may 
represent a more efficient method  to measure the fraction of binaries
in a cluster for several reasons:
\begin{itemize} 
\item{availability of a large number (thousands) of stars makes it a
  statistically roboust method;}
\item{it is cheep in terms of observational time: two filters are
  enough for detecting binaries and repeated measuremets are not needed.} 
\item{it is sensitive to binaries with any orbital period and
  inclination}
\end{itemize}
This approach have been used by other groups 
( see Sollima et al. 2007 and references within) to study the
population of binaries 
in GGCs. The relative small number of clusters that have been
analized is consequence of the intrinsic difficulties of the method:
\begin{itemize} 
\item{high photometric quality is required;}
\item{in some cases, the differential reddening spreads the MS and
  makes it more difficult to isolate the binary sequence;}
\item{an accurate analysis of photometric errors as well as a correct
  estimate of field contamination are necessary to
  disantangle real binaries from 
   bad photometry and field stars. }
\end{itemize}
In this paper, we analyse the catalogs obtained by Anderson
et al. (2008) from HST ACS/WFC data. We exploited both the
homegeneity of this dataset, and the high photometric accuracy of the
measures to derive the fraction of binaries in the central regions of
fifty GGCs.       
\section{Outliers in the Color Magnitude Diagram}
Binaries that are able to survive in the dense enviroment of a
globular cluster are so close that even the Hubble Space Telescope
(HST) is not able to separate their single components.
For this reason, light coming from each star will combine, and the
binary system will appear as a single pointlike source.

In this paper we will take advantage from this instrumental limit to
search for binaries by analising their peculiar position in the CMD.

In the most general case, if we consider two stars in a binary system
and indicate with $m_{1}$, $m_{2}$, $F_{1}$ and $F_{2}$ their
respective magnitudes and fluxes, a simple algebric count demonstrate
that the binary will appear as a single object with a magnitude:
\begin{center}
$m_{bin}=m_{1}-2.5~log(1+\frac{F_{1}}{F_{2}})$
\end{center}
In the case of a binary formed by two MS stars (MS-MS), fluxes
are related to stellar masses ($M_{1}$, $M_{2}$), and its luminosity
 depends on the mass ratio $q=M_{2}/M_{1}$ (in the following we will
 assume $M_{2}<M_{1}$, $q<1$). 
The binaries formed by an equal mass pair form a sequence parallel to MS,
and $\sim ~ 0.75$ magnitudes brighter.
When the masses of the two components are different, the binary will
appear redder and brighter than the MS and it will be located in a CMD region
on the red side of the MSRL.

An  obious consequence  of this  analysis  is that  our capability  in
detecting binaries  depends meanly by  the photometric quality  of the
data. Binaries with large mass  ratios have a large distance from MSRL
and are relatively easy to be  detected. On the contrary, a small mass
ratio pushes binaries near the  MSRL and makes it hard to separate them from
single MS stars.  Moreover, the poorer photometry of faint stars limits
the luminosity (mass) range where they can be detected and studied.
\section{Method}
The limited photometric precision makes  it impossible to
measure the overall  population of binaries even in  a small region of
clusters.   For this  reason,  in this  paper,  we do  not pretend  to
measure the  global fraction of binaries in a cluster, but will limit our
study  to particular  subsamples  of  them. Each  group  is formed  by
objects that share all the  same properties in terms of luminosity and
mass ratio.

We isolated three samples of  high mass ratio binaries (defined as the
binary systems with, $q>$ 0.5, 0.6 and 0.7) and separately studied the
properties  of each  group. In  addition  we derived  also the  global
fraction of binaries.

We performed our study in the magnitude range 
$3.75 < \Delta(I_{F814W}) < 0.75$ below the main sequence turn off. The
extremes of this interval will be indicated with $I_{bright}$ and
$I_{faint}$. 
\subsection {High $q$ binary fraction}
\begin{figure}
   \centering
   \includegraphics[width=7 cm]{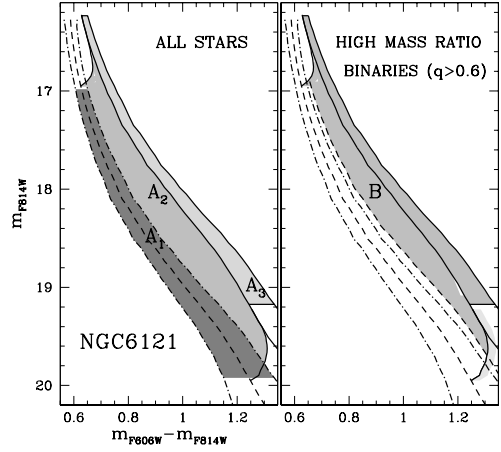}
      \caption{ Grey areas are the regions of the NGC~6121 CMD adopted
       to select all the 
       (single and binary) cluster stars (right) and the
       candidate binaries with q$>$0.6 (left) in a range of 3
       $I_{F814W}$ magnitudes. 
       }
         \label{f1}
   \end{figure}

In order to measure the fraction of high $q$ binaries, we divided the
CMD in two regions (see Fig.\ref{f1}):
 
A region ($A$) that includes all  the MS single stars and the binaries
with a  primary star with  $I_{bright}>I>I_{faint}$.  It is  formed by
three subregions. The  first one ($A_{1}$) includes all  the MS single
stars and MS-MS binaries with small mass ratios; it is limited
by dashed  lines in  Fig.\ref{f1} and corresponds  to the  CMD portion
with a  color distance from the  MS ridge line smaller  than three times
the  MS  dispersion; the  second  ($A_{2}$)  includes  all the  binary
candidates with  high mass ratios.  In Fig.\ref{f1} it  corresponds to
the CMD portion on the red side of the $A_{1}$ region and is delimited
by:
\begin{itemize}
\item {the  track formed by a  binary system with a  primary star with
 $I=I_{bright}$ and a mass ratio ranging from 0 to 1 on the top;}
\item {the corresponding track for a binary system with a primary star
of mass $I_{faint}$ on the bottom;}
\item {the ridge line for an equal mass binary system on the red side.}
\end{itemize}
The third region  ($A_{3}$) contains all the binaries  with $q \sim 1$
that are shifted  by photometric errors to the right  of an equal mass
binaries fiducial line. It is adiajent to the region $A_{2}$ and it is
limitated by the ridge line for an equal mass binary system shifted in
color by three times the main sequence photometric dispersion on the left side.

The second region $B$ is defined as the portion of the region ($A$) on
the red side of the track  formed by a binary star with $q=q_{tr}$ and
it  includes  all   the  binaries  formed  by  a   primary  star  with
$I_{bright}>I>I_{faint}$ and  a mass ratios greater  than a threashold
value ($q_{tr}$).  In this work  we separately studied the  samples of
binaries with mass ratios larger than $q_{tr}$=0.5, 0.6 and 0.7.

Unfortunatly, regions $A$ and $B$  are populated by field stars, while
chance superposition  of two  unrelated stars (apperent  binaries) may
reproduce the behaviour of a genuine binary system.

To  estimate  the   quantity  of  background/foreground  objects  that
casually  overlap the  cluster  CMD  we used  the  galactic models  of
Girardi et al. (2008) (with  the ecception of seven clusters, where we
could isolate field  stars  through  proper  motions).   The  fraction  of
apperent binaries  has been quantified by  performing artificial star
tests.

We further  applied a  the tecnique described in Milone  et al.  2008 to
correct the  spread in color  caused by differential  reddening and/or
spatially dependent zero point photometric errors.
   
In order  to measure the fraction  of high mass ratio  $q$ binaries, we
started  by deriving  the observed  numbers  of stars  in regions  $A$
($N_{OBS}^{A}$)  and  $B$  ($N_{OBS}^{B}$).   Then  we  evaluated  the
corresponding   values   of   arificial   stars   ($N_{ART}^{A}$   and
$N_{ART}^{B}$) and  field stars ($N_{FIELD}^{A}$  and $N_{FIELD}^{B}$)
The correct numbers of real, field and artificial stars are calculated
as $N=\sum{1/c_{i}}$, where $c_{i}$ is the completeness.

High mass binary fraction is calculated as 
\begin{center}
 $f_{bin}=\frac {N_{OBS}^{B}-N_{FIELD}^{B}} {N_{OBS}^{A}-N_{FIELD}^{A}} - \frac {N_{ART}^{B}}{N_{ART}^{A}} $
\end{center}
\subsection {The global binary fraction}
In the  case of  MS-MS binaries with  small mass ratios,  the distance
from  the MSRL  is comparable  to the  size of  photometric  errors in
color. Therefore  these objects  appear mixed to  single stars,  and a
more  sophisticate statistical  analysis is  required to  derive their
contribution to the global fraction of binaries.

In order to estimate the  global fraction of binary systems we adopted
 a statistical method which is based on the comparison of the observed
 data  with more  then  10,000 simulated  CMDs  enriched by  different
 fractions of binaries  with a given $f(q)$. Details  of this tecnique
 are  outside the  purpose of  this work  and will  be expleined  in a
 farcoming  paper (Milone et  al. 2008).   We want  emphasize here
 that  the adopted  statistical approach  is very  sensitive  to small
 inaccuracies and  the results  that we present  are to  be considered
 just as approximate estimates of the real fraction of binaries
\section{Results}
We  investigated if  the fraction  of binaries  depends on  the radial
distance from the cluster center. To this aim we divided the ACS field
into four concentric annuli, each containing roughly the same number
of stars, and measured the fraction of binaries in each of them.

In (at least) the $\sim$ 60 \%  of the 37 GGCs with good photometry in
the innermost  regions, binaries are more  centrally concentrated than
single  MS  stars.  In   the  other  objects,  their  distribution  is
consistent with a flat distribution, probably, as a consequence of the
small radial coverage of our field of view.

In the following, we present preliminary results that involve binaries selected
from different cluster regions:
\begin{itemize}
\item{inside the core radius ($I_{CORE}$ sample);}
\item{between the core and the half-mass radius ($HM$ sample);}
\item{outside the half-mass radius ($O_{HM}$ sample).}
\end{itemize} 
Even if data used in this paper
are homogeneous as they came from the same observing facility
(ACS/HST) and have been reduced adopting the same tecniques,
their photometric quality (and completeness) may vary from one cluster to the
other, mainly, as a consequence of the different stellar densities.

For this reason, it was possible to include in the $I_{CORE}$ sample
only 35 out 50 GGCs. In addition, the limited ACS field of view reduced
the number of GGCs with $HM$ and $O_{HM}$ samples to 46 and 29 respectively.

   \begin{figure}
   \centering
   \includegraphics[width=7. cm]{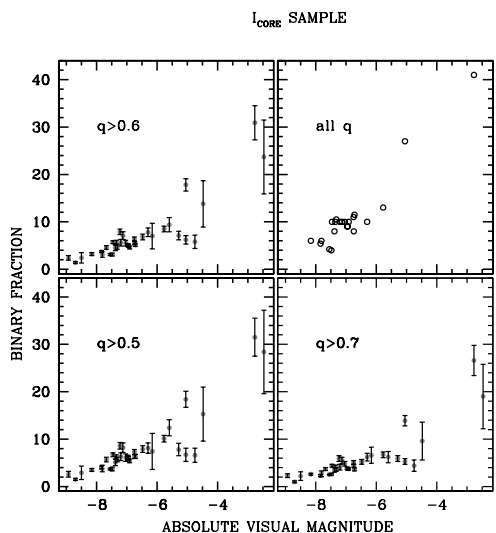}
      \caption{ Fraction of binaries with $q>$ 0.5,
       0.6 and 0.7 and global fraction of binaries in the core as a
       function of the host GGCs luminosity. 
       }
         \label{f2}
   \end{figure}
   \begin{figure}
   \centering
   \includegraphics[width=7. cm]{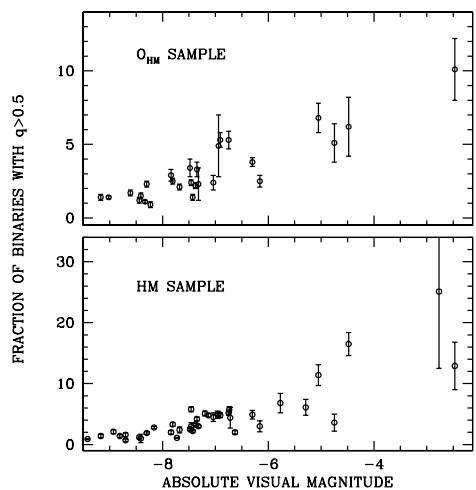}
      \caption{ Fraction of binaries with $q>0.5$ for the $O_{HM}$
       (top) and $HM$ samples (bottom)
       }
         \label{f3}
   \end{figure}

We explored possible relations between the fraction of binaries and the
main parameters of their host GGCs  
absolute visual magnitude, metallicity, collisional parameter, core
and half mass relaxation time, central density and concentration (from
Harris et al. 1996, 2003).

We found an highly significant anticorrelation between the binary
fraction and the total cluster luminosity with clusters with fainter
absolute luminosity have higher 
binary fractions. This anti-correlation is shown in
Fig.\ref{f2} for the $I_{CORE}$ sample and in Figs.\ref{f3} and
for the $HM$ (bottom panel) and $O_{HM}$ (upper panel) samples. 

A  similar anticorrelation  has  been  found by  Piotto  et al  (2004) and
Moretti et al. (2008) between  the frequency of blue stragglers and the
luminosity of  the cluster.  It  is very tempting to  connect BSS and binaries
populations in GGCs: the BSS frequency can be in fact related to the
evolution of the binary fraction due to encounters as described by
Davies et al. (2004). We found only a marginal
correlation between the binary fraction and the cluster collisional parameter
   \begin{figure}
   \centering
   \includegraphics[width=6. cm]{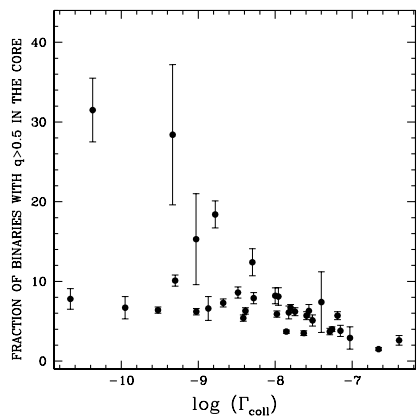}
      \caption{ Fraction of binaries with $q>0.5$ in the core as a
       function of the collisional parameter (Davies et al. 2004)
       }
         \label{f4}
   \end{figure}
\bibliographystyle{aa}

\end{document}